\documentclass[conference]{IEEEtran}
\IEEEoverridecommandlockouts
\usepackage{cite}
\usepackage{amsmath,amssymb,amsfonts}
\usepackage{algorithmic}
\usepackage{array}
\usepackage{graphicx}
\usepackage{textcomp}
\usepackage{xcolor}
\usepackage{hyperref}
\usepackage{url}
\usepackage{booktabs}
\usepackage{multirow}
\usepackage{pifont}

\usepackage{setspace}
\usepackage{parskip}
\usepackage{hyperref}
\usepackage{url}
\usepackage{cite}
\usepackage{booktabs}
\usepackage{array}
\usepackage{xcolor}
\usepackage{graphicx}
\usepackage{amsmath}
\usepackage{amssymb}
\usepackage{enumitem}
\usepackage{titlesec}
\usepackage{fancyhdr}
\usepackage{float}
\usepackage{caption}
\usepackage{subcaption}
\usepackage[ruled,vlined,linesnumbered]{algorithm2e}
\usepackage{listings}
\usepackage{tabularx}
\usepackage{multirow}
\usepackage{mdframed}

\def\BibTeX{{\rm B\kern-.05em{\sc i\kern-.025em b}\kern-.08em T\kern-.1667em\lower.7ex\hbox{E}\kern-.125emX}}

\SetAlgoLined
\SetKwProg{Contract}{Contract}{ is}{end}
\SetKwProg{Func}{Function}{:}{end}
\SetKwProg{Modifier}{Modifier}{:}{end}
\SetKwFunction{FhasRole}{hasRole}
\SetKwFunction{FgrantRole}{grantRole}
\SetKwFunction{FrevokeRole}{revokeRole}
\SetKwFunction{FcreateExam}{createExam}
\SetKwFunction{FenrollStudent}{enrollStudent}
\SetKwFunction{FtransitionState}{transitionState}
\SetKwFunction{FregisterHash}{registerHash}
\SetKwFunction{FverifyHash}{verifyHash}
\SetKwFunction{FsubmitMarks}{submitMarks}
\SetKwFunction{FsubmitScrutiny}{submitScrutiny}
\SetKwFunction{FfinalizeResult}{finalizeResult}
\SetKwFunction{FgetAuditTrail}{getAuditTrail}

\begin{document}

\title{ParikkhaChain: Blockchain-Based Result Processing and 
Privacy-Preserving Academic Record Management for the 
Complete Examination Lifecycle }

\author{\IEEEauthorblockN{Rabib Jahin Ibn Momin}\IEEEauthorblockA{\textit{Department of CSE, BUET}\\Dhaka, Bangladesh}
\and
\IEEEauthorblockN{Ahmed Mahir Sultan Rumi }\IEEEauthorblockA{\textit{Department of CSE, BUET}\\Dhaka, Bangladesh}
\and
\IEEEauthorblockN{Dr. Rezwana Reaz}\IEEEauthorblockA{\textit{Department of CSE, BUET}\\Dhaka, Bangladesh}}

\maketitle

\begin{abstract}
Academic examination systems worldwide continue to rely on centralised,
opaque record-keeping that is often vulnerable to credential forgery, result
tampering, examiner bias, and the absence of transparent re-evaluation
pathways.  Existing blockchain-based approaches in education focus
predominantly on post-hoc certificate storage or online-only examination
portals, leaving the complete onsite examination lifecycle---from conducting exams through scrutiny---largely unaddressed.  This paper proposes
\textbf{ParikkhaChain}, a blockchain-based framework that covers the entire
examination lifecycle of an onsite examination system with three distinguishing contributions: (i)
\emph{anonymous script evaluation} through cryptographic hashing of answer
scripts before examiner access, thereby eliminating identity-based bias; (ii)
a \emph{transparent evaluation and scrutiny workflow} backed by an immutable on-chain audit
trail that records every mark submission and grade revision; and (iii) Inclusion of privacy-preserving verification using zero-knowledge proofs and off-chain 
storage mechanisms.  The system is architected around four Solidity smart contracts
deployed on the Ethereum blockchain. The proposed architecture is the first initiative to our knowledge to support physical examination process, anonymous marking, and
re-evaluation transparency. We successfully simulate full exam cycles of an onsite exam to grade-sheet generation using a working prototype on a large scale of 100 courses and hundreds of teachers and students. The experimental results show that the system can manage online examinations of hundreds of courses, students and faculties efficiently with great throughput, little storage and transaction cost. Our codebase is available in open source form at \url{https://github.com/AhmedRumi/CSE6608-ParikkhaChain}.

\end{abstract}


\section{Introduction}

The integrity of academic examinations is foundational to educational equity: grades shape university admissions, scholarship awards, and
professional licensing.  Yet the processes that produce those grades---script
registration, examiner allocation, marking, result tabulation, and
re-evaluation---are overwhelmingly managed through centralised, paper-based
or monolithic digital systems, specially in developing countries like Bangladesh that offer little transparency, no anonymity,
and no tamper-evident history of changes.  Credential fraud, result
manipulation, and systemic examiner bias remain pervasive concerns across
educational systems globally \cite{chinnasamy2025blockchain,shirke2025trustexaminator}.
 
Blockchain technology, characterised by its decentralised distributed
ledger, cryptographic immutability, and programmable smart-contract layer,
has attracted growing research interest as a mechanism for securing
educational records.  Early investigations focused on issuing verifiable
digital certificates \cite{hybrid2024maritime} or managing final transcripts
\cite{chinnasamy2025blockchain}.  A second wave of research applied blockchain
specifically to online examination portals, demonstrating improvements in
authentication, question-paper integrity, and result transparency
\cite{sattar2023advanced,shirke2025trustexaminator}.  Despite these
advances, critical gaps persist: no system simultaneously addresses the
\emph{full lifecycle} of a physical examination, provides on-chain
anonymisation of answer scripts during marking, or records a transparent
audit trail of grade revisions through a formal scrutiny process.
 
This paper presents \textbf{ParikkhaChain}, a blockchain-based framework
designed to address these gaps.  The system manages the entire examination
lifecycle---exam creation, student enrolment, script registration, anonymous
evaluation, marks submission, scrutiny, and result finalisation---using a
suite of interoperable Ethereum smart contracts. The key design principles and contributions of the project are the following.
 
\begin{itemize}

    \item \textbf{Anonymous Script Evaluation.}  Answer scripts are uploaded
    in encrypted form and registered on-chain by a cryptographic hash that is
    decoupled from the student's identity.  Examiners receive scripts without
    any identity of the student.
 
    \item \textbf{Immutable Audit Trail.}  Every mark submission and every
    subsequent grade revision is permanently recorded on the distributed
    ledger.
 
    \item \textbf{Role-Based Access Control (RBAC).}  A dedicated RBAC smart
    contract governs four actor roles---Admin, Student, Examiner, and
    Scrutinizer---and all other contracts query it before permitting any
    state-changing action.
 
    \item \textbf{Selective Zero-Knowledge Proof Verification.}  For
    third-party credential checks (e.g.\ by employers or universities),
    students can request a ZKP that proves certain criteria, thereby preserving privacy while
    enabling trust.
 
    \item \textbf{Off-chain Storage with On-chain Anchoring.}  Large binary
    artefacts (answer scripts, grade sheets) are stored off-chain.
\end{itemize}

\section{Literature Review}
\label{sec:litrev}
% ══════════════════════════════════════════════════════════════════════════════
 
The literature on blockchain in education can be partitioned into two broad
clusters in our context: (i) \emph{academic credential and document management} systems that
primarily target final-outcome certificates and transcripts, and (ii)
\emph{online examination systems} that use blockchain to secure the delivery
and grading of digital tests.  
 
% ─────────────────────────────────────────────────────────────────────────────
\subsection{Academic Credential and Document Management Systems}
% ─────────────────────────────────────────────────────────────────────────────
 
A significant body of work on credential security focuses primarily on protecting final academic records rather than the underlying examination process. For instance, Chinnasamy et al.~\cite{chinnasamy2025blockchain} proposed a blockchain-based framework augmented with machine learning techniques for detecting malicious activity and classifying documents, though it is limited to post-issuance records. Similarly, Pandey et al.~\cite{pandey2023idms} introduced an AI-driven Intelligent Document Management System (IDMS) to improve document handling efficiency, but it relies on centralized infrastructure and lacks blockchain-based guarantees. A hybrid approach by~\cite{hybrid2024maritime} integrates Hyperledger Fabric with IoT and QR codes to enable secure certificate issuance and verification, yet it remains confined to the certification stage. Collectively, these studies emphasize securing the output of examinations while leaving critical upstream processes—such as marking, re-evaluation, and grade handling—outside the trust boundary.
% ─────────────────────────────────────────────────────────────────────────────
\subsection{Blockchain-Based Online Examination Systems}
% ─────────────────────────────────────────────────────────────────────────────
The second strand of research applies blockchain technology directly to the examination process, primarily in online settings. Sattar et al.~\cite{sattar2023advanced} proposed a blockchain-based framework for securing online examinations, using encrypted and hash-linked data storage to prevent issues such as question leakage and answer tampering, though it is limited to fully digital workflows. Similarly, Shirke et al.~\cite{shirke2025trustexaminator} introduced TrustExaminator, a comprehensive system leveraging the Ethereum blockchain, smart contracts, and facial recognition for secure and automated examination management; however, it remains focused on online exams and does not address anonymised evaluation of physical scripts or privacy-preserving re-evaluation processes.
 
% ─────────────────────────────────────────────────────────────────────────────
\subsection{Identified Gaps and Motivation for ParikkhaChain}
% ─────────────────────────────────────────────────────────────────────────────
 
The existing literature highlights several persistent limitations in existing systems. Prior work predominantly focuses on securing final academic outputs—such as grades and certificates—while neglecting the underlying examination processes that generate them \cite{chinnasamy2025blockchain,hybrid2024maritime,pandey2023idms}. Moreover, most blockchain-based solutions are designed for fully digital environments and do not support physical, paper-based examinations that remain prevalent in many educational contexts \cite{sattar2023advanced,shirke2025trustexaminator}. Additionally, these systems fail to incorporate mechanisms for anonymised evaluation, leaving grading vulnerable to potential bias, and they do not model scrutiny (re-evaluation) workflows or maintain transparent, immutable audit trails for grade changes. Collectively, these gaps indicate that critical aspects of fairness, accountability, and end-to-end trust in examination systems remain insufficiently addressed.

% ParikkhaChain directly addresses all four gaps by covering the
% examination lifecycle from exam creation to grade finalisation on-chain, supporting physical answer-script upload and encrypted IPFS storage, 
% anonymising scripts via the HashRegistry smart contract before
% examiner access, and implementing a Scrutinizer role with a ResultAudit
% contract that mandates and persists reasons for every grade change.  The
% selective use of ZKP for third-party verification further extends the
% privacy guarantees beyond what any reviewed system provides, making
% ParikkhaChain a novel and comprehensive contribution to the intersection of
% blockchain technology and examination management.

\section{System Design and Methodology}
\label{sec:design}
% ══════════════════════════════════════════════════════════════════════════════

\begin{figure}[!h]
    \centering
    \includegraphics[width=0.5\textwidth]{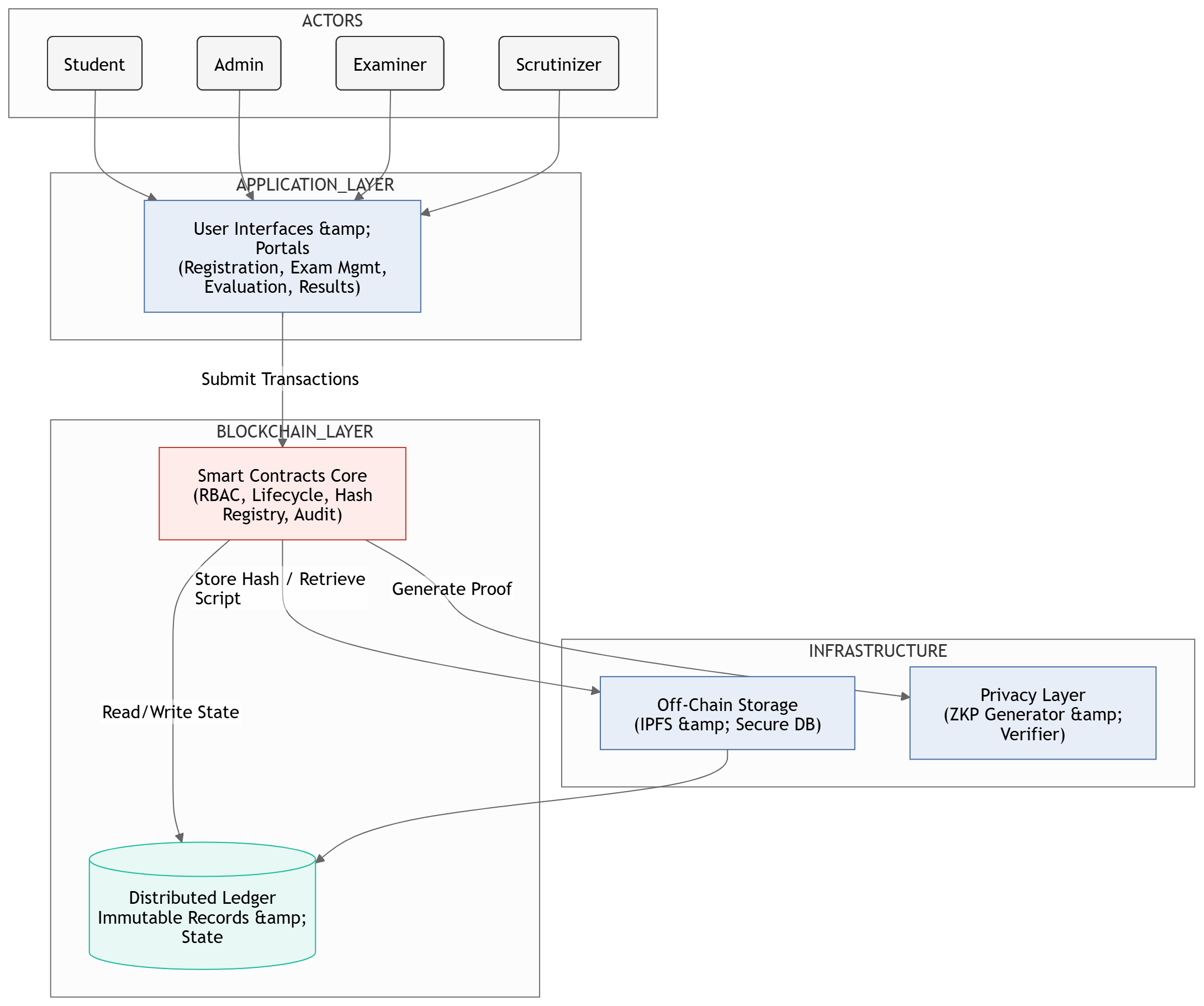}
    \caption{High-level architecture of ParikkhaChain showing the four-layer
    design: actors, application portals, blockchain smart contracts, and
    infrastructure (IPFS and ZKP).}
    \label{fig:hla}
\end{figure}

ParikkhaChain is organised into four conceptual layers: the \emph{Actor
Layer}, the \emph{Application Layer}, the \emph{Blockchain Layer}, and the
\emph{Infrastructure Layer}. Figure~\ref{fig:hla} presents the high-level
architecture illustrating how these layers interact. The system is run through four smart contracts named RBAC, HashRegistry, ResultAudit, and ExamLifeCycle Contract. 

Actors can interact with the system through role-specific application portals. Portals
translate those actions into signed Ethereum transactions that invoke
smart-contract functions on the blockchain. The smart-contract core
reads and writes state to the distributed ledger, whilst delegating large
binary data to the off-chain IPFS storage and credential proofs to the
ZKP privacy layer.

% ─────────────────────────────────────────────────────────────────────────────
\subsection{System Actors}
% ─────────────────────────────────────────────────────────────────────────────

The system defines four distinct actors, each with a well-scoped set of
privileges:

\begin{itemize}[]
    \item Admin --- The exam authority. Responsible for creating
    exams, enrolling students, assigning examiners, and triggering lifecycle
    state transitions. All administrative actions are verified against the
    RBAC contract before execution.
    \item Student --- Examination candidates. A student attends the exam being held onsite (not part of the system), and queries their results and grade sheets through
    the Result Verification portal.
    \item Examiner --- The evaluators. Examiners can access encrypted, identity-free
    answer scripts via the Evaluation Portal for their assigned courses, assign marks, and submit them 
    to the ResultAudit contract. Crucially, the examiner never sees the
    student's identifying information during this phase.
    \item Scrutinizer --- The re-evaluators. Reviews submitted marks, may revise the grade, and must supply a written
    justification string that is permanently recorded on-chain.
\end{itemize}

\subsection{Application Layer}

The application layer  exposes six portals that
map directly onto the examination lifecycle:

\begin{enumerate}
    \item \textbf{Registration Portal} --- Student enrolment and identity
    creation. Generates a register transaction invoking the RBAC contract.
    \item \textbf{Exam Management} --- Admin portal for creating exams and
    managing state. Issues a Create Exam Transaction.
    \item \textbf{Script Upload Interface} --- Accepts scanned/digitised answer
    scripts, encrypts them, stores them on IPFS, and writes the resulting
    content hash via an Upload Hash Transaction.
    \item \textbf{Result Verification} --- Allows students to query their
    published results using a Query Transaction to the ResultAudit contract.
    \item \textbf{Evaluation Portal} --- Allows examiners and scrutinizers
    to retrieve de-identified scripts from IPFS (via the stored hash) and
    submit marks. Issues Submit Marks Transactions.
    \item \textbf{Grade Sheet Generator} --- Produces PDF/CSV grade sheets
    from finalised on-chain results, storing the generated document in the
    Secure Database.
\end{enumerate}

The detailed interaction between actors, portals, and smart contracts is
shown in Figure~\ref{fig:bclayer}.

\begin{figure}[!h]
    \centering
    \includegraphics[width=0.5\textwidth]{Figures/fig_2.png}
    \caption{Blockchain and application layer flow diagram showing how
    each actor's portal communicates with the four smart contracts
    (RBAC, ExamLifecycle, HashRegistry, ResultAudit) and the distributed ledger.}
    \label{fig:bclayer}
\end{figure}

\subsection{Smart Contracts}

The four main smart contracts are the heart of the blockchain system. They are described in this section.
\subsubsection{Role-Based Access Control (RBAC Contract)}
\label{subsec:rbac}
% ─────────────────────────────────────────────────────────────────────────────

The RBAC Contract is the security foundation of the entire system. Every
other smart contract in the system queries RBAC Contract before
executing any role-specific function; an operation that fails the role check
is immediately reverted. The contract maps Ethereum addresses to roles and
supports granting, revoking, and querying role memberships.

Four roles are defined as \texttt{ADMIN}, \texttt{EXAMINER}, \texttt{SCRUTINIZER}, and \texttt{STUDENT}.
A mapping \texttt{roleRegistry[address] $\rightarrow$ Role} stores the
current role assignment for each registered address. An additional mapping
tracks enrolment status of students per examination, used as a precondition
for script registration.

Algorithm~\ref{alg:rbac} presents the pseudocode for the core RBAC
operations.

\begin{algorithm}[!h]
\caption{RBAC Contract --- Role Management}
\label{alg:rbac}
\DontPrintSemicolon
\SetAlgoLined

\textbf{State Variables:}\\
\quad $roleRegistry$: mapping(address $\rightarrow$ Role) \tcp*{Role enum: NONE, ADMIN, EXAMINER, SCRUTINIZER, STUDENT}
\quad $enrollmentRegistry$: mapping(examId $\times$ address $\rightarrow$ bool)\;
\quad $deployer$: address \tcp*{Set at deployment; has bootstrap ADMIN privilege}

\BlankLine
\Func{\FgrantRole{$targetAddr$, $role$}}{
    \If{$msg.sender \neq deployer$ \textbf{and} $roleRegistry[msg.sender] \neq \text{ADMIN}$}{
        \textbf{revert} ``Caller is not Admin''\;
    }
    \If{$targetAddr = \text{address}(0)$}{
        \textbf{revert} ``Invalid address''\;
    }
    $roleRegistry[targetAddr] \leftarrow role$\;
    \textbf{emit} $RoleGranted(targetAddr, role, msg.sender)$\;
}

\BlankLine
\Func{\FrevokeRole{$targetAddr$}}{
    \If{$roleRegistry[msg.sender] \neq \text{ADMIN}$}{
        \textbf{revert} ``Caller is not Admin''\;
    }
    $roleRegistry[targetAddr] \leftarrow \text{NONE}$\;
    \textbf{emit} $RoleRevoked(targetAddr, msg.sender)$\;
}

\BlankLine
\Func{\FhasRole{$addr$, $expectedRole$}}{
    \Return $roleRegistry[addr] = expectedRole$\;
}

\BlankLine
\Func{$enrollStudent$($examId$, $studentAddr$)}{
    \If{\textbf{not} \FhasRole{$msg.sender$, $\text{ADMIN}$}}{
        \textbf{revert} ``Only Admin can enroll students''\;
    }
    \If{$roleRegistry[studentAddr] \neq \text{STUDENT}$}{
        \textbf{revert} ``Address is not a registered Student''\;
    }
    $enrollmentRegistry[examId][studentAddr] \leftarrow \text{true}$\;
    \textbf{emit} $StudentEnrolled(examId, studentAddr)$\;
}

\BlankLine
\Func{$isStudentEnrolled$($examId$, $studentAddr$)}{
    \Return $enrollmentRegistry[examId][studentAddr]$\;
}
\end{algorithm}

% ─────────────────────────────────────────────────────────────────────────────
\subsubsection{Examination Lifecycle Contract}
\label{subsec:lifecycle}
% ─────────────────────────────────────────────────────────────────────────────

The ExamLifecycle Contract governs the creation of examination records and
enforces strictly ordered state transitions. Each examination instance
progresses through five states: $CREATED \rightarrow ACTIVE \rightarrow SUBMITTED \rightarrow SCRUTINIZED \rightarrow COMPLETED.$

No contract function is permitted to skip a state. This linear state machine
ensures, for example, that an examiner cannot submit marks before the exam is
active, and that a scrutinizer cannot review results before the initial marks
have been submitted.

The contract also validates student enrolment (via a cross-contract call to
RBAC Contract) before accepting script hash registrations, and
exposes a \texttt{getExamState()} view function that other contracts query as
a precondition guard.

\subsubsection{Hash Registry and Anonymous Script Evaluation}
\label{subsec:hashregistry}
% ─────────────────────────────────────────────────────────────────────────────

The HashRegistry Contract is the mechanism that implements anonymous
evaluation. When a student's physical answer script (the topsheet) is scanned and digitized,
the application layer encrypts the file and uploads it to IPFS, receiving
back a content-addressed CID (Content Identifier). A \emph{Script ID}
(e.g., \texttt{TS\_7a8f\ldots}) is generated independently of the student's
identity and is the only identifier recorded on-chain. The mapping between
Script ID and student identity is held off-chain by the Admin and is only
revealed after all marks have been finalised.

The contract maintains a simple but powerful invariant: once a hash is
registered for a Script ID, it is immutable. Any attempt to re-register
the same Script ID is rejected, preventing tampering with the physical
record after evaluation begins.

\begin{figure*}[!h]
    \centering
    \includegraphics[width=0.9\textwidth, height=200pt]{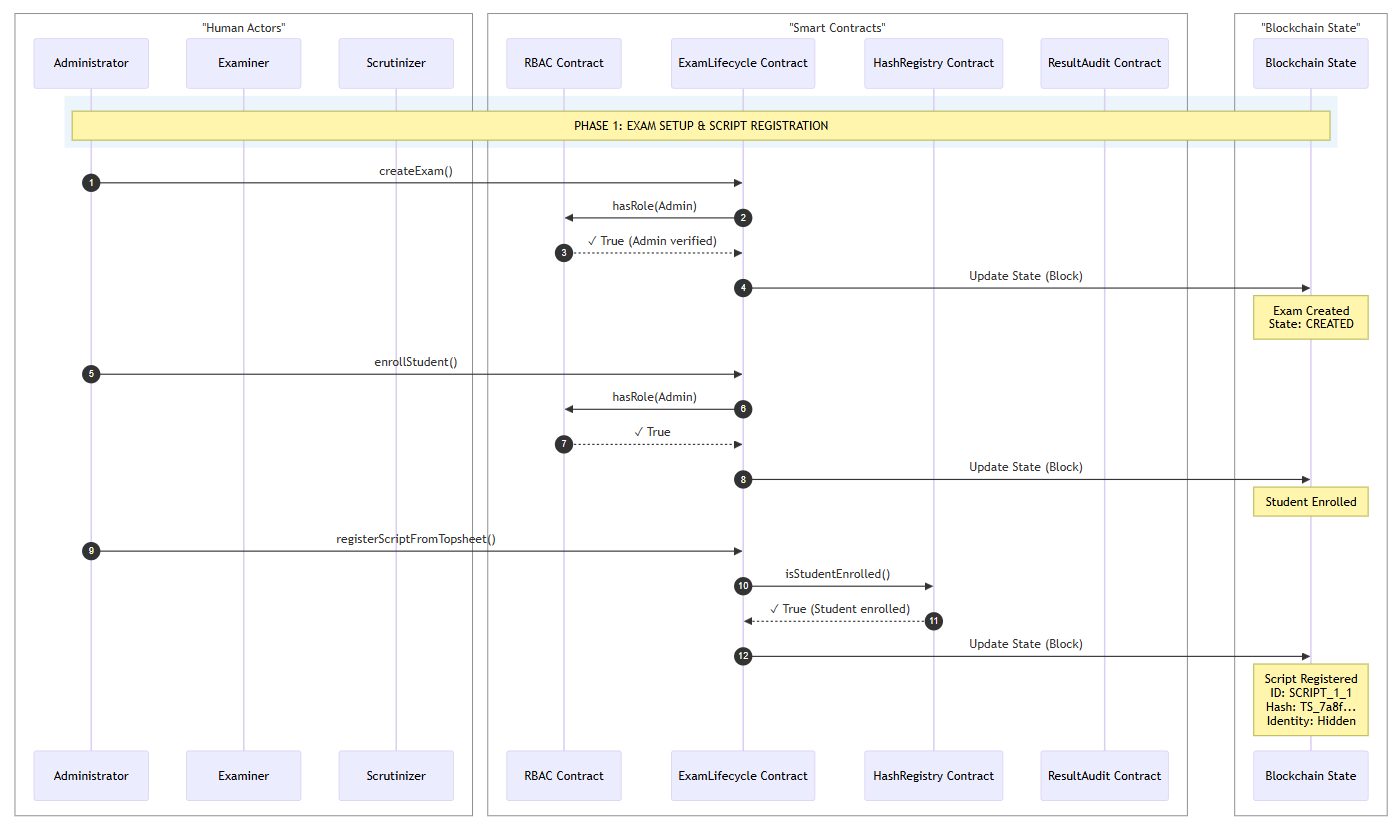}
    \caption{Exam Setup \& Script Registration}
    \label{fig:exam}
\end{figure*}

Exam Setup and Script Registration process involves RBAC, Exam Lifecycle and HashRegistry contracts. This process is shown in Fig. ~\ref{fig:exam}.

% ─────────────────────────────────────────────────────────────────────────────
\subsubsection{Result Audit Contract}
\label{subsec:resultaudit}
% ─────────────────────────────────────────────────────────────────────────────

\begin{figure*}[!h]
    \centering
    \includegraphics[width=0.9\textwidth, height=200pt]{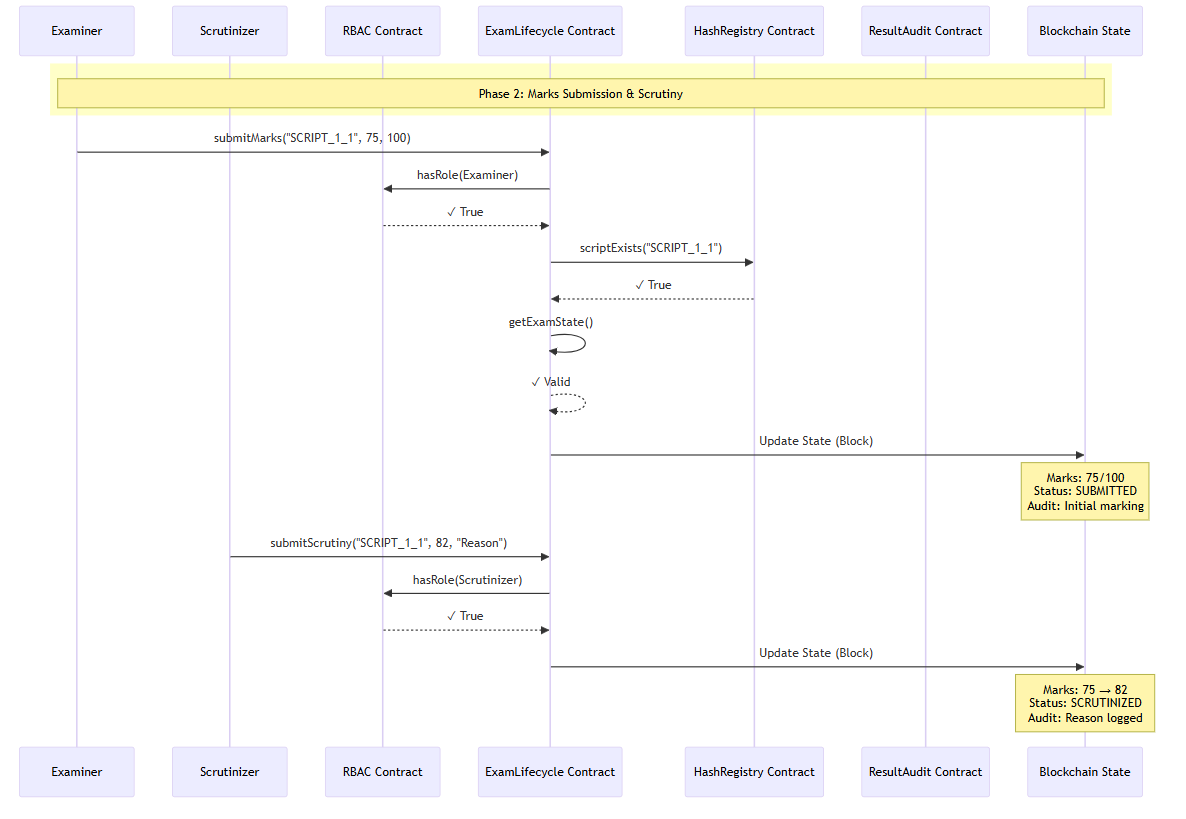}
    \caption{Marks Subsmission \& Scrutiny}
    \label{fig:marks}
\end{figure*}

The ResultAudit Contract is the most operationally complex component of the
system. It manages the full marks lifecycle: initial submission by the
examiner, optional revision by the scrutinizer (with mandatory justification),
and final result publication. Every state change produces an immutable event
log on the blockchain, constituting the tamper-evident audit trail. The marks submission and scrutiny process involves interactions among all the contracts, as illustrated in Fig.~\ref{fig:marks}.

% ─────────────────────────────────────────────────────────────────────────────
\subsection{Off-Chain Storage Architecture}
\label{sec:offchain}
% ─────────────────────────────────────────────────────────────────────────────

\begin{figure}[!h]
    \centering
    \includegraphics[width=0.48\textwidth]{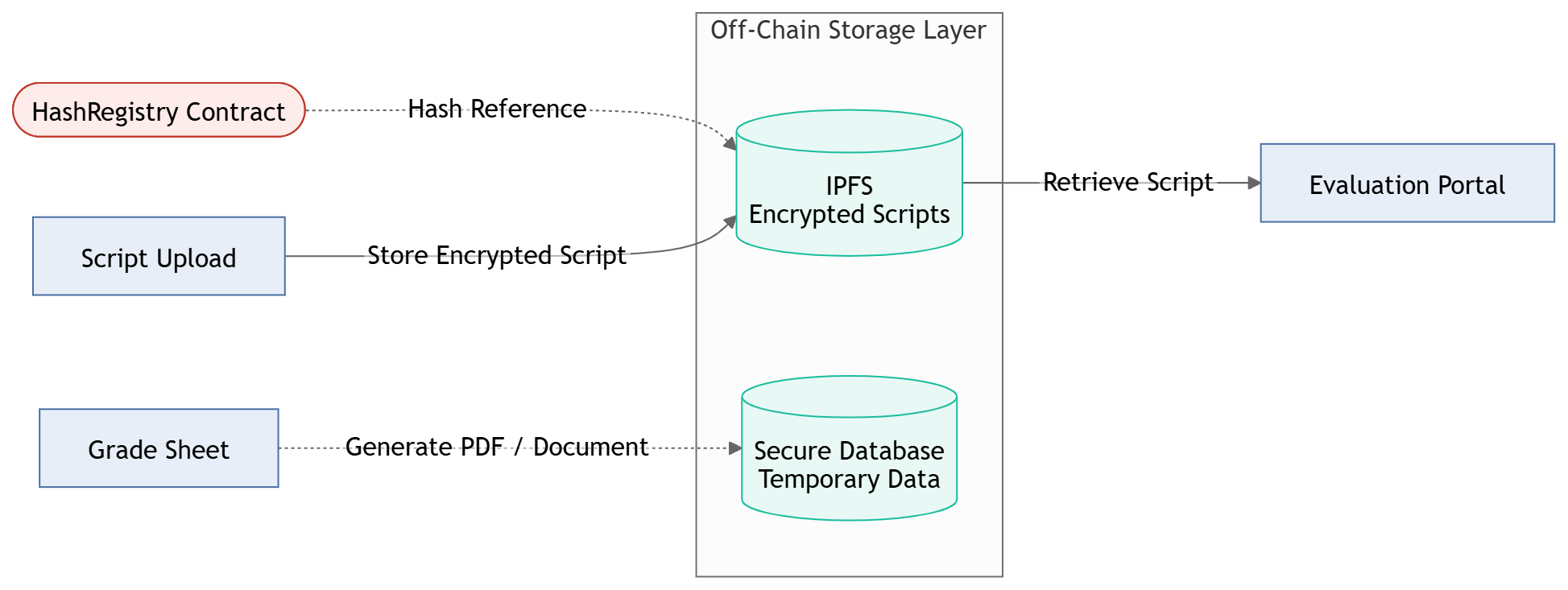}
    \caption{Off-chain storage architecture. Encrypted scripts are persisted
    on IPFS; the HashRegistry contract holds only the cryptographic hash
    reference. Grade-sheet documents (PDF/CSV) are generated into a
    temporary secure database.}
    \label{fig:offchain}
\end{figure}

Storing raw answer-scripts directly on a blockchain ledger is
impractical: Ethereum's storage cost grows linearly with data size, and
answer scripts---even as compressed PDFs---may be tens of megabytes each. Usually the scanning of the top sheet should suffice. The answer scripts can be checked offline which do not contain any identity of the student.  The topsheet and also the grade sheets and certificates of the student are stored offchain. 
ParikkhaChain therefore adopts a \emph{hybrid} storage strategy illustrated
in Figure~\ref{fig:offchain}.

The off-chain storage layer comprises two components:

\paragraph{IPFS (InterPlanetary File System) for Encrypted Scripts.}
A student's topsheet of the answer script can be uploaded via the Script Upload Interface,
the application layer performs the following steps:
\begin{enumerate}
    \item Encrypt the script file using a symmetric key held by the Admin
    (e.g., AES-256-GCM).
    \item Register the topsheet of the script which will return a content-addressed CID.
    \item Compute \texttt{keccak256(CID)} to produce a fixed-length 32-byte
    hash.
    \item Call HashRegistry Contract.\texttt{registerHash(scriptId, hash)}
    to anchor the reference on-chain immutably.
\end{enumerate}

When an Examiner or Scrutinizer requests a script via the Evaluation Portal,
the portal retrieves the hash from the blockchain, fetches the corresponding
ciphertext from IPFS, decrypts it using the Admin-controlled key, and
presents the plaintext script---still without any student identifying
metadata.

\paragraph{Secure Database for Temporary Data.}
Grade sheets generated as PDF or CSV documents are stored in a secure,
access-controlled relational database as temporary data. These documents
are produced from finalised on-chain results and are made available to
students through the Result Verification portal. Because grade sheets are
derivative outputs (their source of truth is the immutable ledger), no
long-term off-chain storage guarantee is required for them.

% ─────────────────────────────────────────────────────────────────────────────
\subsection{Privacy Layer and Zero-Knowledge Proofs}
\label{sec:zkp}
\begin{figure}[!h]
    \centering
    \includegraphics[width=0.5\textwidth]{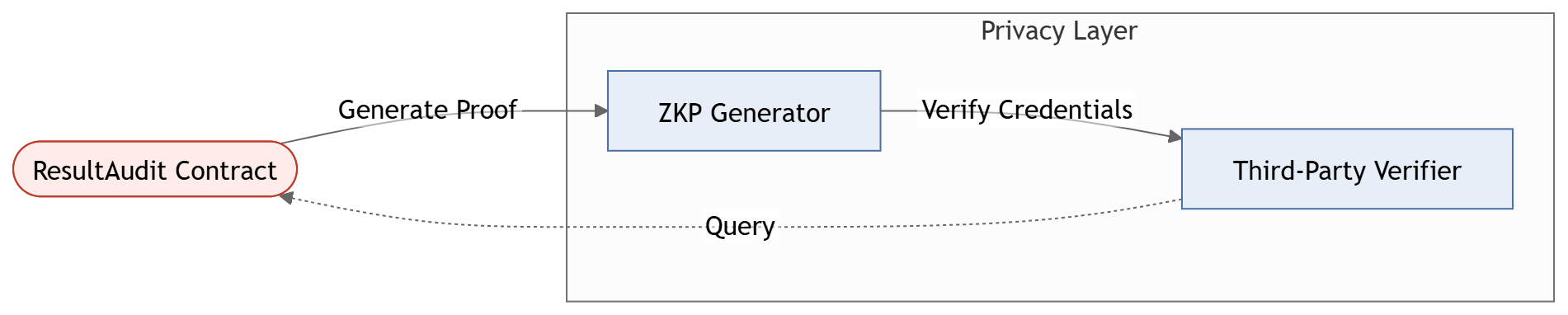}
    \caption{Privacy layer architecture. The ResultAudit contract generates a
    proof request; the ZKP Generator constructs a validity proof; the
    Third-Party Verifier validates the proof without accessing the underlying
    grade data.}
    \label{fig:privacy}
\end{figure}

% ─────────────────────────────────────────────────────────────────────────────
Zero-Knowledge Proofs (ZKPs) in blockchain are cryptographic methods allowing one party (prover) to prove to another (verifier) that a statement is valid—such as having sufficient funds or authorized identity—without revealing the underlying private data. We use selective Zero-Knowledge Proof (ZKP) verification for some operations in the system. Here, the prover is student and the verifier is an employer or any other third-party verifier. A recurring challenge in academic credential systems is the tension between
\emph{verifiability} and \emph{privacy}. For example, a student may wish to prove to a
third-party employer that they passed an examination without revealing their
exact score. ParikkhaChain addresses this through \emph{selective} Zero-Knowledge
Proof (ZKP) verification, as illustrated in Figure~\ref{fig:privacy}. A student can generate a proof using ZKP Generator which can be presented to the third party without verifying the whole information of the student. The verifier validates the proof mathematically using the ZKP
    Verifier component. No query to the blockchain ledger is needed, and
    no grade data is revealed. 

\section{Experimental Setup}
\label{sec:setup}
% ══════════════════════════════════════════════════════════════════════════════

This section describes the implementation stack, development environment and the metrics used for the ParikkhaChain prototype. The primary
goal of the prototype is a \emph{full-cycle demonstration}: to exercise every
phase of the exam cycle end-to-end on a local or test-net
Ethereum environment using mock examination data, and to evaluate the system's
functional correctness and workflow integrity.

% ─────────────────────────────────────────────────────────────────────────────

\subsection{Blockchain Layer}

The smart contracts are implemented in \textbf{Solidity} (target compiler
\texttt{$\geq$0.8.x}) chosen because of its strong static typing, and the maturity of its tooling
ecosystem. All four contracts (RBAC Contract,
Exam Lifecycle Contract, Hash Registry Contract,
ResultAudit Contract) are authored and compiled using
\textbf{Remix IDE}. We worked on both the Ethereum Virtual
Machine (EVM) and local machine. Remix
provides an integrated Solidity compiler, a JavaScript EVM for rapid local
testing, and a one-click deployment interface to public testnets.

\subsection{Mock Data, Automation Scripts and Offchain Storage}

For the prototype in this project, we have not implemented the frontend or user portals for the system. Rather, we directly call the functions of the smart contracts to simulate the exam cycle. To automate the process, we generate large amount of mock data. The mock data contains a set of courses, students, and teachers, each course containing the list of registered students and assigned evaluators and scrutinizers. 

For the offchian storage, we just use our local machine storage in this project.

% \subsubsection{Development and Test Environment}

\begin{table}[!h]
\centering
\caption{Development environment configuration}
\label{tab:devenv}
\renewcommand{\arraystretch}{1.3}
\begin{tabular}{p{3cm} p{5cm}}
\toprule
\textbf{Component} & \textbf{Configuration} \\
\midrule
Smart Contract Language & Solidity \texttt{$\geq$0.8.20} \\
IDE & Remix IDE (Web App) \\
Local Ethereum Node & Remix JavaScript VM  \\
Backend Language & Python 3 \\
Blockchain Library & Web3.py v6.x \\
Hash Library & \texttt{hashlib} + \texttt{pysha3} (keccak256) \\
IPFS (prototype) & Mock local file server \\
ZKP (prototype) & Commit-Reveal scheme \\
Grade Sheet Generation & Python \texttt{reportlab} (PDF), \texttt{csv} module \\
Version Control & Git \\
\bottomrule
\end{tabular}
\end{table}

% ─────────────────────────────────────────────────────────────────────────────

\subsection{Evaluation Metrics}
% ─────────────────────────────────────────────────────────────────────────────

The prototype will be evaluated against the following metrics:

\begin{itemize}
    \item \textbf{Throughput:} Number of transactions processed per minute. 
    \item \textbf{Gas Cost:} Ethereum gas cost (in ETH and dollars) for
    contract deployment and transaction processing, measured on
    the Remix IDE.
    \item \textbf{Storage:} Storage needed for the smart contracts and the data generated in the system.
\end{itemize}

% \newpage

% ─────────────────────────────────────────────────────────────────────────────

% \newpage
% ══════════════════════════════════════════════════════════════════════════════
\section{Results}
\label{sec:results}
% ══════════════════════════════════════════════════════════════════════════════

The ParikkhaChain prototype was deployed and evaluated on Remix ID Desktop App using our local machine. We used Python automation scripts to simulate the exam scenarios.
Performance was measured across three exam scenarios of increasing scale ---
\emph{Small}, \emph{Medium}, and \emph{Large} --- to assess throughput, gas
cost, and on-chain storage overhead. Gas pricing was referenced at 0.044 gwei and ETH at \$2154 USD. Storage overhead is
estimated using a gas-based estimation (Estimates bytes written to storage during workflow execution. Approximately 35\% of smart contract gas is consumed by SSTORE
operations — an empirical figure consistent with Ethereum gas profiling literature  ~\cite{wood2014ethereum}.
The \textbf{Large} scenario is the primary benchmark, representing a
realistic departmental workload of the CSE Department, BUET.

% ─────────────────────────────────────────────────────────────────────────────
\subsection{Scenario Overview}

We had 5, 25, and 100 course exams with 43, 605, and 4,439 scripts registered in the small, medium, and large
scenarios, respectively. All of the 261, 2,747, and 18,930 transactions in these scenarios completed with a \textbf{0\% failure rate}, confirming
the correctness of the smart-contract logic and the robustness of the
automation scripts across widely varying workload sizes.

% ─────────────────────────────────────────────────────────────────────────────
\subsection{Large-Scale Scenario: BUET CSE Departmental Workload}
\label{sec:large}
% ─────────────────────────────────────────────────────────────────────────────

The Large scenario was designed to mirror the realistic examination workload
of the CSE Department, BUET: 100 exams covering approximately 4,439 student
scripts across all courses and semesters in a given academic term.
Table~\ref{tab:large} presents the complete performance profile for this
scenario.

\begin{table*}[!h]
\centering
\caption{Full performance metrics --- Large scenario (BUET CSE workload)}
\label{tab:large}
\renewcommand{\arraystretch}{1.35}
\begin{tabularx}{\textwidth}{p{5cm} p{3.5cm} p{2cm} X}
\toprule
\textbf{Metric} & \textbf{Value} & \textbf{Accuracy} & \textbf{Notes} \\
\midrule
\multicolumn{4}{l}{\textit{Metric 1 --- Throughput}} \\
\midrule
Total transactions             & 18,930          & Exact     & Incl.\ 4 deployments \\
Workflow transactions          & 18,926          & Exact     & Excl.\ deployments \\
Failed transactions            & 0               & Exact     & 0\% failure rate \\
Elapsed time                   & 9,555 s (2h 39m)& Exact     & Ganache local chain \\
Throughput (total)             & 1.98 tx/s       & Exact     & 118.87 tx/min \\
Throughput (workflow)          & 1.98 tx/s       & Exact     & Excl.\ deployments \\
\midrule
\multicolumn{4}{l}{\textit{Metric 2 --- Transaction Cost}} \\
\midrule
Total gas used (all)           & 4,989,664,028   & Exact     & Incl.\ deployments \\
Workflow gas used              & 4,981,280,046   & Exact     & Excl.\ deployments \\
Avg gas per workflow tx        & 263,197         & Exact     & \\
Workflow cost (ETH)            & 0.22 ETH       & Estimated & @ 0.044 gwei \\
Workflow cost (USD)            & \$472.30    & Estimated & @ ETH = \$2154.93 \\
Avg cost per tx (USD)         & \$15.79         & Estimated & \\
\midrule
\multicolumn{4}{l}{\textit{Metric 3A --- Execution Storage (Gas-based Estimation)}} \\
\midrule
Storage fraction of workflow gas & 35\%          & Estimated & Per EIP-2929 \\
SSTORE operations              & 87,172 ops      & Estimated & $= 35\% \times G / 20{,}000$ \\
Execution storage written      & 2,724.12 KB     & Estimated & $87{,}172 \times 32$ bytes \\
\midrule
\multicolumn{4}{l}{\textit{Gas Breakdown by Contract}} \\
\midrule
ResultAudit Contract   & 2,374,867,868 gas (47.6\%) & Exact & 9,128 tx,  260,173 avg gas/tx \\
HashRegistry Contract  & 2,027,244,993 gas (40.6\%) & Exact & 4,439 tx, 456,689 avg gas/tx \\
ExamLifecycle Contract & 546,600,233 gas (11.0\%)   & Exact & 4,939 tx, 110,670 avg gas/tx \\
RBAC Contract          & 32,566,952 gas (0.7\%)     & Exact & 420 tx, 77,540 avg gas/tx \\
Deployments                         & 8,383,982 gas (0.2\%)      & Exact & 4 contracts \\
\bottomrule
\end{tabularx}
\end{table*}

\paragraph{Throughput.}
The system processed 18,930 transactions in 9,555 seconds (2 hours 39 minutes),
sustaining a throughput of 1.98 tx/s on the local environment. This
throughput is sufficient for examination management workflows, where
transactions are batched by the automation layer rather than submitted
interactively in real time. The throughput is slightly lower than the medium
scenario (2.24 tx/s) because the larger number of \texttt{SSTORE}-heavy
ResultAudit Contract and HashRegistry Contract
transactions increases average block processing time at high volume.

\paragraph{Gas Cost.}
The ResultAudit Contract and HashRegistry Contract
together account for 88.2\% of total workflow gas, reflecting the high
storage cost of persisting mark records, audit log entries, and script
hashes for 4,439 scripts. The RBAC Contract consumes only
0.7\% of workflow gas, confirming that the role-check guard pattern adds
negligible overhead. The approximate total workflow cost is \$472.30, which is quite feasible for a full exam cycle of a semester for a department. 

\paragraph{On-Chain Storage.}
Using the gas-based estimation method, the workflow generated
an estimated 87,172 \texttt{SSTORE} operations, writing approximately
\textbf{2.66 MB} of execution state to the distributed ledger for 4,439
scripts. The fixed bytecode footprint of 36.07 KB (across all four
contracts) is invariant across scenarios and well within Ethereum's EIP-170
per-contract size limit of 24,576 bytes. Thus, the memory requirement is quite low indeed. 

% ─────────────────────────────────────────────────────────────────────────────
\subsection{Scalability Analysis}
\label{sec:scalability}
% ─────────────────────────────────────────────────────────────────────────────

\begin{figure}[!h]
    \centering
    \includegraphics[width=0.5\textwidth]{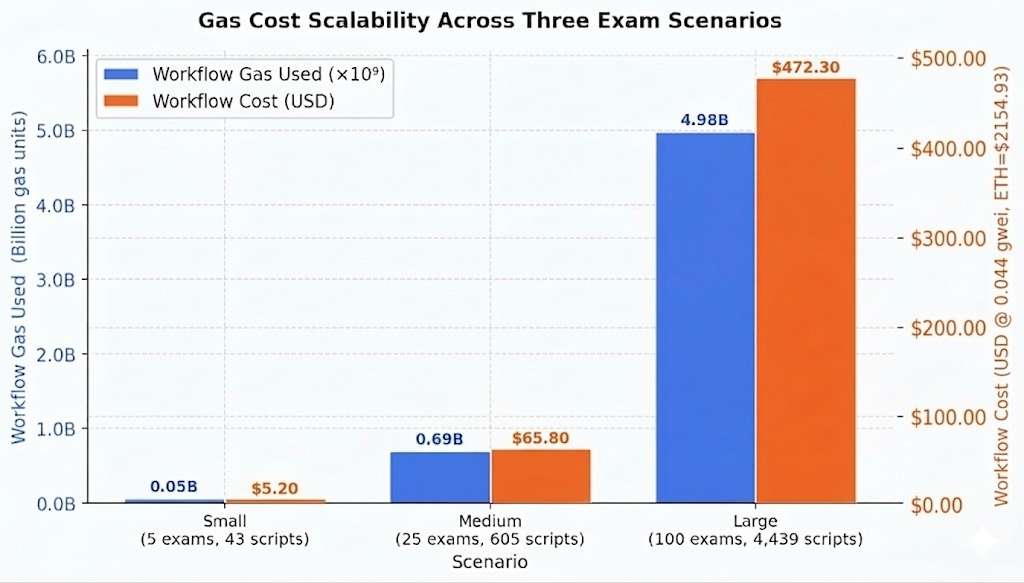}
    \caption{ Gas cost scalability across three exam scenarios. Left axis
    (blue bars): total workflow gas consumed in billions of gas units. Right
    axis (orange bars): equivalent USD cost at 0.044 gwei gas price and
    ETH = \$2154.93. Both metrics scale near-linearly with the number of
    scripts registered.}
    \label{fig:gas_scale}
\end{figure}

\begin{figure}[!h]
    \centering
    \includegraphics[width=0.5\textwidth]{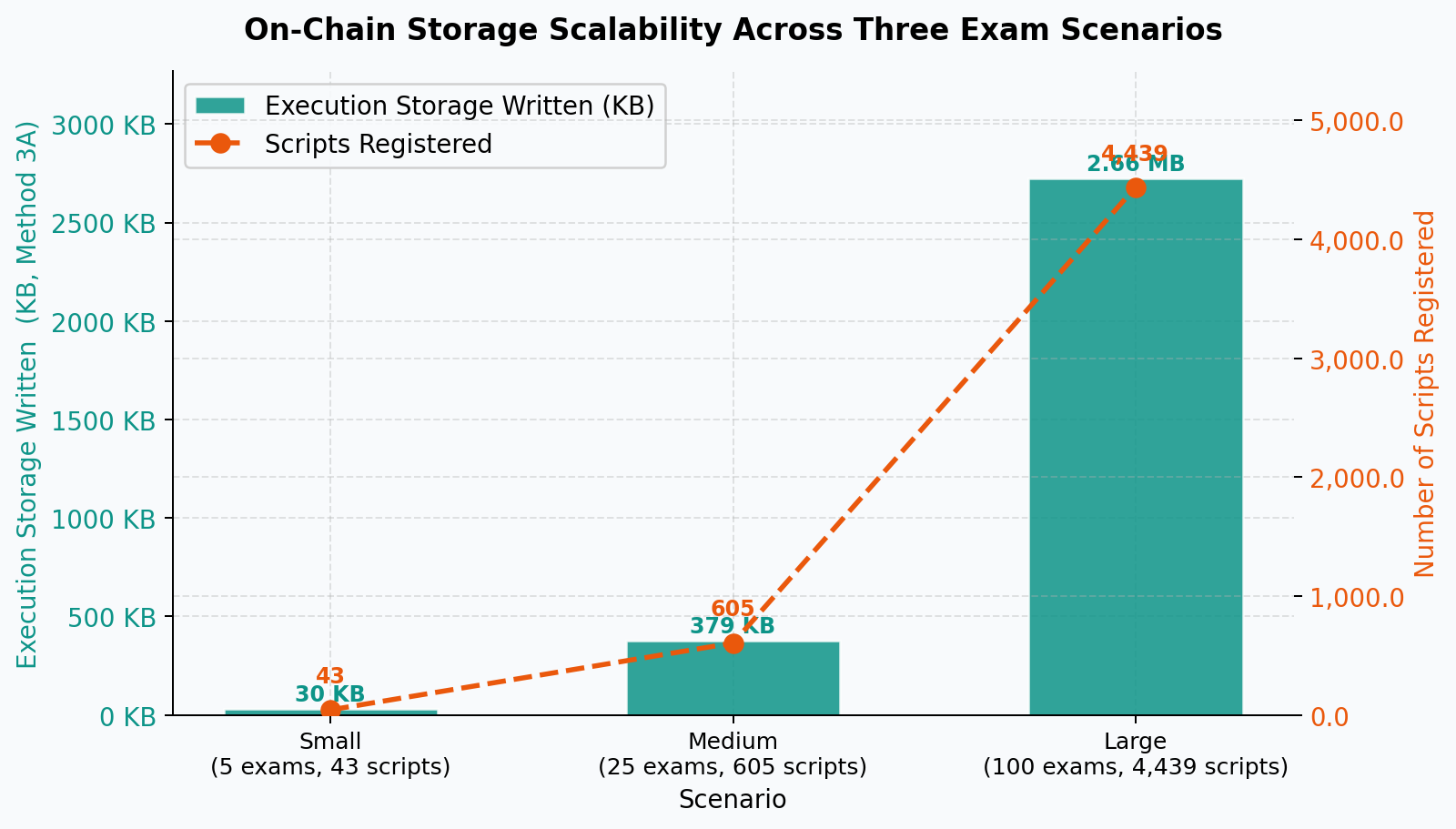}
    \caption{On-chain execution storage scalability across three exam scenarios
    (gas-based estimation). Bars show estimated storage written to
    the ledger in KB; the dashed orange line tracks the number of scripts
    registered. Storage growth closely follows script count, confirming
    linear scaling behaviour.}
    \label{fig:storage_scale}
\end{figure}

Figure~\ref{fig:gas_scale} and Figure~\ref{fig:storage_scale} present the
scalability of gas cost and on-chain storage overhead respectively as the
examination workload grows from the Small to the Large scenario.

As presented in Figures \ref{fig:gas_scale} and \ref{fig:storage_scale}, we observe that the system exhibits near-linear growth in both gas consumption and storage as the number of scripts increases, with minor sub-linear behavior at smaller scales due to amortized fixed setup costs. As workload grows, per-script costs dominate, making resource usage predictable and proportional to script count. Additionally, throughput remains stable across all scales, indicating that performance is not impacted by increased workload and is primarily constrained by block time rather than system complexity. Overall, the system proves to be quite scalable and can handle even thousands of exams, millions of students and teachers with proper resources.

\subsection{Zero Knowledge Proof}
This scheme demonstrates a Zero-Knowledge Proof (ZKP) protocol operating through four phases among three actors — Admin, Student, and a Third Party. In Phase 1, the Admin retrieves finalized marks and generates a random salt, computing a commitment hash off-chain using keccak256 over the scaled CGPA, marks, credits, and salt, then storing only this hash on-chain while keeping actual academic data private. In Phase 2, the Third Party posts eligibility criteria on-chain — including minimum CGPA, grade thresholds, credit requirements, and pass conditions — making them publicly visible and tamper-proof. In Phase 3, the Student proves their eligibility by submitting their actual marks, CGPA, and salt to the contract, which recomputes and verifies the hash against the stored commitment, evaluates all criteria, and records only a boolean outcome without ever storing raw marks on-chain. Finally, in Phase 4, the Third Party queries the contract and receives solely a true/false eligibility result — learning nothing about the underlying academic details — which is the core ZKP property: proving knowledge of a fact without revealing the fact itself. Detailed screenshots of the whole process are given in our github repository.
\section{Conclusion and Future Work}
\label{sec:conclusion}
% ══════════════════════════════════════════════════════════════════════════════

This paper presented \textbf{ParikkhaChain}, a blockchain-based system for
managing the complete lifecycle of onsite academic examinations with anonymous
evaluation, transparent scrutiny, and privacy-preserving result verification across three examination scenarios of increasing scale. The system achieved a \textbf{0\% transaction failure rate} across all
18,930 transactions in the largest scenario, sustained a stable throughput
of approximately \textbf{1.98 tx/s}, and successfully registered 4,439
scripts and 100 exams representing the realistic workload of the CSE
Department, BUET. Gas cost and on-chain storage scaled near-linearly with
script count: the Large scenario consumed 4.98 billion gas units and wrote
an estimated 2.66 MB of execution state to the ledger, confirming predictable
and plannable resource growth. The HashRegistry Contract
and ResultAudit Contract together account for 88.2\% of
workflow gas, reflecting the storage intensity of anonymous script anchoring
and immutable audit-log maintenance --- the two most novel contributions of
the system. We showed that the system that efficiently scales with the largest scenario requiring an approximate amount of \$472 and minimal storage.  

The prototype can be further developed in many directions, including multi-semester transcript management, degree certificate issuance on blockchain, advanced ZKP for granular verification, Support for diverse exam formats, an automated testing suite, cross-institutional result verification, Mobile and web applications and so on. It is also a first step towards developing a national education blockchain consortium for national-level examinations. 

\newpage
% ══════════════════════════════════════════════════════════════════════════════
%  BIBLIOGRAPHY
% ══════════════════════════════════════════════════════════════════════════════

\end{document}